\documentclass[prd,11pt]{revtex4}
\usepackage{amsmath}
\usepackage{amssymb}
\usepackage{graphicx}

\newcommand\new[1]{#1}
\newcommand{\Section}[1]{\section{#1} \setcounter{subsection}{1}}

\date \today
\begin{document}
%\title{Casimir effect in linear optical backgrounds}
%\title{Casimir effect in bi-refringent linear optical backgrounds}
\title{Quantization of general linear electrodynamics}
\author{Sergio Rivera}
\author{Frederic P. Schuller}
\address{Albert Einstein Institute,\\ Max Planck Institute for Gravitational Physics, Am M\"uhlenberg 1, 14476 Potsdam, Germany}
\begin{abstract}
General linear electrodynamics allow for an arbitrary linear constitutive relation between the field strength two-form and induction two-form density if crucial hyperbolicity and energy conditions are satisfied, which render the theory predictive and physically interpretable. Taking into account the higher-order polynomial dispersion relation and associated causal structure of general linear electrodynamics, we carefully develop its Hamiltonian formulation from first principles. Canonical quantization of the resulting constrained system then results in a quantum vacuum which is sensitive to the constitutive tensor of the classical theory. As an application we calculate the Casimir effect in a bi-refringent linear optical medium.
\end{abstract}
\maketitle

\section{Introduction}
Classical electromagnetism can be formulated on much more general optical backgrounds than the familiar ones described in terms of Lorentzian manifolds. From the point of view of electrodynamics, this is because one merely needs a constitutive law that links the electromagnetic field strength two-form $F$ with the induction two-form density $H$, 
and thus closes the relations
\begin{equation}
 (d F)_{\alpha \beta \gamma} = 0, \qquad (d H)_{\alpha \beta \gamma} = \epsilon_{\alpha \beta \gamma \delta} j^\delta\,,
\end{equation}
which any electromagnetic theory featuring charge conservation and no magnetic monopoles in four dimensions must satisfy in the presence of a current vector field density $j$. This point has been made most prominently and lucidly by \cite{hehl2003}. Even if one restricts attention to linear constitutive laws \footnote{Born-Infeld theory presents the most prominent example of electrodynamics with a non-linear constitutive relation between the induction and field strength.}, 
new{the resulting electrodynamic theories will generically feature birefringence, meaning that distinguished polarizations of light will travel at different speeds.Now the most general action for an electromagnetic gauge potential that results in a linear  constitutive law, and which we will carefully quantize in this paper, is}
\begin{equation}
\label{areametricaction}
 S[A,G]=-\frac{1}{8}\int \, dx^4 \omega_G \left[ F_{\alpha \beta} F_{ \gamma \delta}\, G^{\alpha \beta \gamma \delta} + j^\alpha A_\alpha\right]\,,
\end{equation}
where $G$ is a smooth covariant rank four tensor field with the symmetries $G_{\alpha \beta \gamma \delta} = G_{\gamma \delta \alpha \beta}$ and $G_{\alpha \beta \gamma \delta} = -G_{\beta \alpha \gamma \delta}$, and which is invertible in the sense that there is a smooth contravariant tensor field $G^{\alpha \beta \gamma \delta}$
so that $G^{\alpha \beta \rho \sigma} G_{\rho \sigma \gamma \delta} = 2(\delta ^\alpha _\gamma \delta ^\beta _\delta  - \delta^\alpha_\delta \delta ^\beta _\gamma )$ and there is a well-defined volume form $\omega_G$ for such area metric tensors \cite{punzi2007}. The birefringence of such general linear electrodynamics is encoded in its dispersion relation, or equivalently the causal structure, of the associated field equations \cite{raetzel2010}. This dispersion relation is known \cite{Hehl2002} to be of higher polynomial order, and indeed the central challenge faced in this paper is to properly deal with this fact, both in the classical and quantum analysis.
%\new{The causal structure of the dynamics (\ref{areametricaction}) is determined by the area metric tensor $G$, as was shown in the research article \cite{raetzel2010}, whose results are of great relevance for the present paper and will be concisely reviewed in section in section (\ref{causal_structure}). 
The importance of understanding Maxwell theory on such general linear backgrounds is that the latter comprehensively describe all linear optical backgrounds ranging from fundamental spacetime geometries beyond Lorentzian geometry \cite{schuller2005,schuller2006,punzi2009,schuller2010,raetzel2010} over the effective spacetime structure seen by photons to first order quantum corrections in a curved Lorentzian spacetime \cite{drummond1980} to all non-dissipative linear optical media available in the laboratory \cite{schuller2010}.

The present article develops the canonical quantization of these most general linear electrodynamics from first principles, and arrives at an explicit calculation of the quantum vacuum of the theory. We show that the related Casimir effect detects deviations from a non-birefringent background with an amplification which in principle is limited only by technological constraints.
Arriving at these results requires special care when obtaining the Hamiltonian formulation of the classical theory that precedes the actual quantization. While quite generally the Dirac-Bergmann quantization procedure of course also applies to these gauge field dynamics, the key issue is the question of which hypersurfaces provide viable initial data surfaces on which the canonical phase space variables can be defined and evolved by the Hamiltonian. It is precisely this question that makes the problem of quantization of the dynamics (\ref{areametricaction}) so subtle, and requires the conceptually robust understanding of its causal structure developed in \cite{raetzel2010} and concisely summarized in section \ref{causal_structure}. Only when using the insights gained there, can the formulation of the Hamiltonian picture in section \ref{sec_hamiltonian} and the canonical quantization in section \ref{sec_quantization} proceed as usual, based on the derivation of the Dirac brackets in section \ref{sec_dirac} and the diagonalization of the Hamiltonian in section \ref{sec_diagonalization}, which is particularly simple for the area metrics in a neighborhood of Lorentzian metric geometries, as shown in section \ref{sec_classI}. However, having gone through the laborious quantization procedure, one is rewarded in section \ref{sec_casimir} with the said method to measure deviations from a metric-induced background through the Casimir effect in particular, and a demonstration of how to quantize field theories with higher-order polynomial dispersion relations \cite{ling2006,barcelo2007,rinaldi2007,garattini2010,bazo2009,gregg2009,sindoni2008,sindoni2009,sindoni2009-2,chang2008,laemmerzahl2009,laemmerzahl2005,liberati2001,perlick2010,gibbons2007,lukierski1995,glikman2001,girelli2007} in general. 
      
For simplicity, we restrict attention to flat area metric manifolds throughout the paper. Analogous to any other geometric structure on a smooth manifold, an area metric is called flat if there exists a set of charts covering the underlying smooth manifold such that the components of the area metric tensor are constant within each such chart.

\section{Causual structure of linear electrodynamics}
\label{causal_structure}
The Hamiltonian formulation of the dynamics (\ref{areametricaction}), on which the canonical quantization will be built, hinges on several key results of the associated causality theory. Here we summarize the central results of practical importance. For a detailed derivation of these results we refer the reader to \cite{raetzel2010}. A necessary condition for Maxwell theory on a four-dimensional area metric background to be predictive is that the following polynomial \cite{Hehl2002} on covectors $k$,
\begin{equation}
\label{fresnel}
 P(k)=-\frac{1}{24} \epsilon_{\rho \sigma \tau \epsilon} \epsilon_{\mu \nu \omega \vartheta}  G^{\rho \sigma \mu \alpha} G^{\beta \tau \nu \gamma} G^{\delta \epsilon \omega \vartheta }k_\alpha k_\beta k_\gamma k_\delta\,,
\end{equation}
is hyperbolic. This means that there is at least one covector $h$ with $P(h)\neq 0$ such that for every covector $q$ the polynomials 
\begin{equation}
P_{q,h}(\lambda) = P(q + \lambda h)
\end{equation} 
have only real roots $\lambda$, in which case $h$ is said to be a hyperbolic covector with respect to $P$. 
That hyperbolicity is a necessary criterion for a well-posed initial value problem is a central result of the theory of partial differential equations \cite{garding1964,atiyah1970}. For the flat area metric manifolds discussed here, the hyperbolicity of $P$ is even sufficient for the predictivity of the theory \cite{garding1964}. Initial data, given on a hypersurface whose normal covectors are all hyperbolic with respect to $P$, are then uniquely evolved away from the hypersurface. Thus a Hamiltonian formulation of the dynamics, which deals precisely with the evolution between initial data surfaces, must be based on a foliation $\{t,x^a\}$ of the manifold whose leaves $t=\textrm{const}$ are hypersurfaces with hyperbolic conormal.

However, this requirement needs to be sharpened if one requires that the actual initial data can be collected by observers. The definition of observers now hinges on the so-called dual polynomial $P^\#$ \cite{hassett2007}, which for those $P$ that arise from area metrics by virtue of (\ref{fresnel}) and which admit hyperbolic covectors, can be calculated explicitly and takes the deceivingly simple form
\begin{equation} 
 \label{dualfresnel}
 P^\#(X)=-\frac{1}{24}\epsilon^{\rho \sigma \tau \epsilon} \epsilon^{\mu \nu \omega \vartheta} G_{\rho \sigma \mu \alpha} G_{\beta \tau \nu \gamma} G_{\delta \epsilon \omega \vartheta } X^\alpha X^\beta X^\gamma X^\delta\,,
\end{equation}
which sends any tangent vector $X$ to a real number. 
That the dual polynomial $P^\#$ can be calculated analytically at all, and takes such a comparatively simple form, is only due to an interplay of the area metric structure underlying it and the necessary hyperbolicity of $P$.  
While the hyperbolic covectors of $P$ distinguish admissible initial data surfaces, admissible observers are distinguished by their worldline tangent vectors being hyperbolic vectors of $P^\#$. In other words, the very existence of observers restricts the admissible area metric geometries further to those where also $P^\#$ is hyperbolic. 
But exactly this hyperbolicity of $P^\#$ then allows to make a choice of time-orientation, which in turn implies a choice of positive energy.  More precisely, a time-orientation is chosen by picking one connected set of all hyperbolic tangent vectors, a so-called hyperbolicity cone $C^\#$, out of the several such connected components defined by $P^\#$.  But then the covectors $q$ for which all future-directed observers measure positive energy, $q(X)>0$ for all $X\in C^\#$, themselves constitute a cone $(C^\#)^+$ in cotangent space, which thus deserves to be called the positive energy cone with respect to the chosen time-orientation.
The latter, in turn, selects the (open and convex) cone $C$ of hyperbolic covectors of $P$ that lie within the positive energy cone $(C^\#)^+$. For technical convenience we require, without loss of generality, that $P$ be positive on all of $C$; indeed, from (\ref{fresnel}) it is clear that this always can be arranged for by switching the overall sign of $G$.  

\begin{figure}
\includegraphics[width=15cm,angle=0]{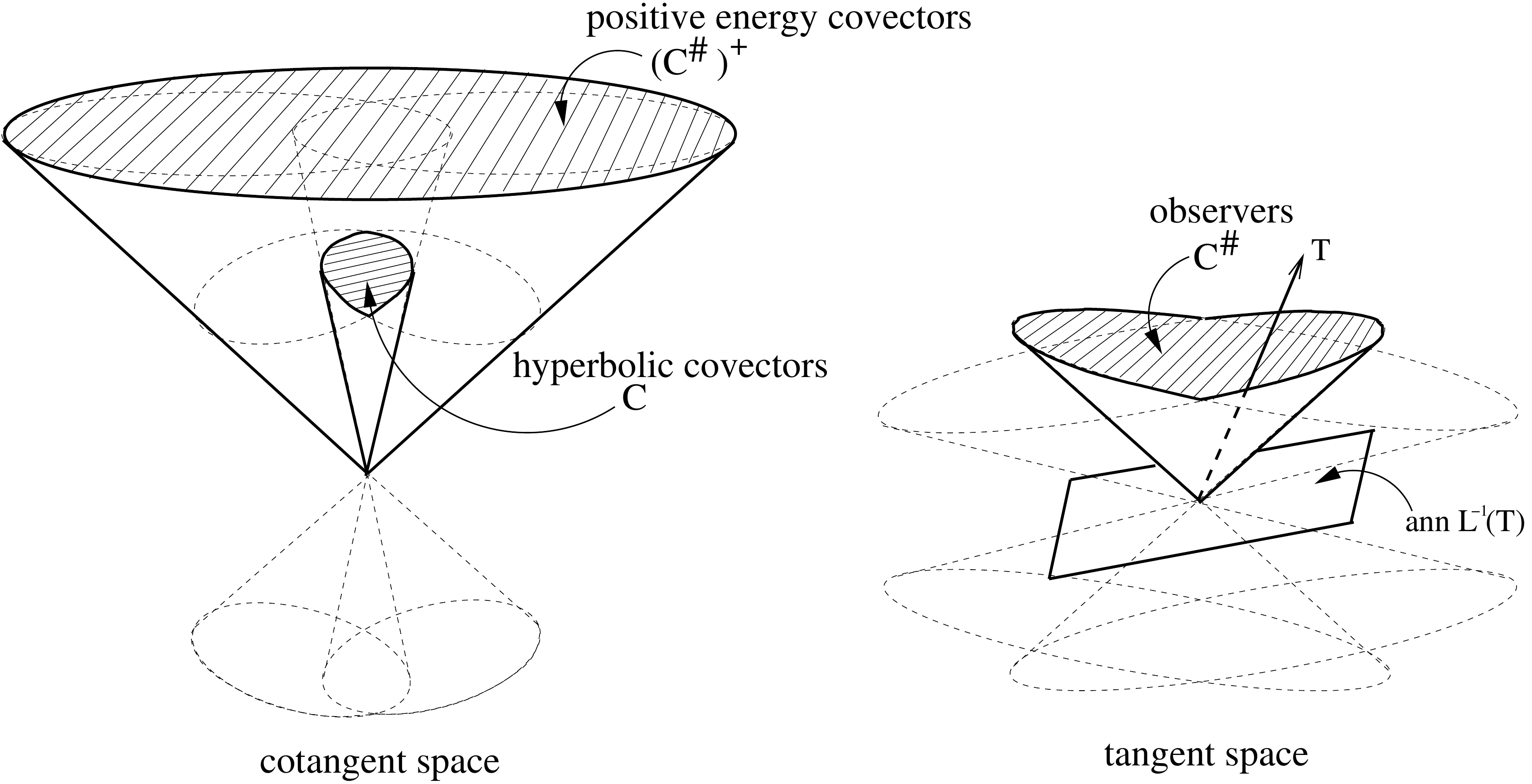}
\caption{\label{fig_cones} Causal structure of a typical bi-hyperbolic and energy-distinguishing polynomial dispersion relation. Dotted surfaces indicate sets of covectors and vectors that are null with respect to the polynomial $P$ and its dual $P^\#$, respectively.  The cone $C$ of hyperbolic covectors and the cone $C^\#$ of observers  both arise as hyperbolicity cones. Purely spatial directions, as seen by an observer with worldline tangent $T$, are those vectors annihilated by the preimage of $T$ under the Legendre map $L$.} 
\end{figure}

Besides the hyperbolicity of $P$ and $P^\#$, one finally needs to require that there exists a time orientation such that any non-zero $P$-null covector lies either in $(C^\#)^+$ or $-(C^\#)^+$. In other words, the energy of any massless momentum is to have a definite sign upon which all observers agree. If and only if this bi-hyperbolicity and energy distinguishing properties are met, it is justified to call the underlying area metric manifold an area metric {\it spacetime}, and we will consider only such. For an illustration in a typical case, see figure \ref{fig_cones}, and for a detailed exposition of these concepts, see \cite{raetzel2010}.
  
The final piece of technology concerns the duality map between covectors and vectors in an area metric spacetime. The map
\begin{equation}
  L: C \to L(C)\,,\qquad L(q) = \frac{P(q,q,q,\cdot)}{P(q,q,q,q)}
\end{equation}
is shown in \cite{raetzel2010} to be a well-defined and invertible Legendre map precisely because $P$ is assumed to be bi-hyperbolic and energy-distinguishing. Spacelike hypersurfaces are meaningfully defined as those having tangent directions that are purely spatial with respect to some observer. More precisely, the spacelike hypersurfaces are those whose conormals lie in $L^{-1}(C^\#)$. But since it can be shown that $L^{-1}(C^\#)$ always lies within $C$, the condition that a hypersurface be spacelike (and thus initial data on it accessible by local observers) further sharpens the condition for a feasible initial data surface for the dynamics (\ref{areametricaction}) we identified before. Thus only a foliation $(t, x^1, x^2, x^3)$ of an area metric spacetime into spacelike hypersurfaces $t=\textrm{const}$ (which then contain all vectors that are annihilated by the covector $L^{-1}(\partial/\partial t)$, see figure \ref{fig_cones}) provides an appropriate temporal-spatial split for the Hamiltonian formulation. For the flat area metric spacetimes considered here, one may further choose the coordinates such that $L^{-1}(\partial/\partial t)$ is conormal to the spacelike hypersurfaces. In other words, one can choose trivial shift and lapse in the flat case.  

The reader may find it helpful to get a feel for these seemingly abstract conditions for the special case where the area metric is induced by a metric $g$ by virtue of $G_{\alpha \beta \gamma \delta} = g_{\alpha \gamma}g_{\beta \delta}-g_{\alpha \delta}g_{\beta \gamma}$. Precisely the same conceptual steps force one then to take the metric $g$ to be of Lorentzian signature (otherwise Maxwell theory would not be well-posed). Since in this metric-induced case $P(k)=(g^{\alpha \beta}k_\alpha k_\beta)$ as usual, we have $L(C)=C^\#$, and thus one recovers the standard Lorentzian notions of obervers and spacelike hypersurfaces. However, the general construction presented before does not justify itself from this reduction to the metric case. The general treatment rather demonstrates the appropriateness and consistency of the standard Lorentzian definitions from a conceptual point of view.

\section{Hamiltonian formulation}
\label{sec_hamiltonian}
With the appropriate foliation $(t,x^1, x^2, x^3)$ of the area metric spacetime with spacelike leaves for constant time $t$ and conormals given by $L^{-1}(\partial/\partial t)$, as constructed in the previous section, we are now in a position to develop the Hamiltonian formulation of the dynamics encoded in the action (\ref{areametricaction}). For a flat area metric spacetime, one can choose coordinates not only such that the area metric has constant components throughout those charts, but also that additionally the components of the volume form $\omega_G$ featuring in the action are numerically identical to those of the totally antisymmetric Levi-Civita symbol $\epsilon_{\alpha \beta \gamma \delta}$ defined by $\epsilon_{0123}=+1$. In such a coordinate system, we obtain the canonical momenta associated with the field variables $(A_0,A_i)$ from the Lagrangian density $\mathcal{L}$ of the action (\ref{areametricaction}) as
\begin{eqnarray}
 \label{canonicalmomenta}
\pi^{0}=\frac{\delta \mathcal L}{\delta (\partial_0 A_0)}&=&0,\\
\nonumber
\pi ^{i}  =  \frac{\delta \mathcal L}{\delta (\partial_0 A_i)} &=& -G^{0i0j}\partial_0 A_j-G^{0ij0}\partial_j A_0-G^{0ijk}\partial_j A_k\,,
\end{eqnarray}
where here, and for the remainder of the paper, latin indices range from 1 to 3, while greek indices continue to range from 0 to 3. 
In the language of the theory for constrained systems \cite{hanson1976,sundermeyer1982}, we thus identify $\phi_1=\pi^{0}\approx 0$ as a primary constraint of the dynamics. Defining the matrix $M_{ij}$ with the property $M_{ij} G^{0i0j}=\delta^k_j$, whose existence is guaranteed if the differential equations coming from (\ref{areametricaction}) are hyperbolic (see appendix \ref{appendixone}), and using (\ref{canonicalmomenta}) to express $\partial_0 A_i$ in terms of the canonical momenta $\pi^{i}$, we find the total Hamiltonian density
\begin{eqnarray}
\label{ext_ham_1}
\mathcal H&=&-\frac{1}{2}M_{js}\pi^j \pi^s-A_0 \partial_j \pi^j-\pi^j M_{ij}G^{0irk}\partial_r A_k\\ \nonumber
& &+\frac{1}{2} G^{ijkr}\partial_i A_j \partial_k A_r-\frac{1}{2} M_{ir}G^{0ijk} G^{0rmn} \partial_m A_n \partial_j A_k + u_1(x) \pi^0(x).
\end{eqnarray}
Following the Dirac-Bergmann algorithm \cite{sundermeyer1982} for obtaining the Hamiltonian formulation of systems with constraints, we now compute the commutator $ \{ \pi ^0,\mathcal H\}$. If this commutator is not zero, we need to impose $ \{ \pi ^0,\mathcal H\}\approx 0$ as a secondary constraint, in order to ensure that the primary constraint $\phi_1\approx 0$ is preserved under time evolution. Indeed, one obtains $ \{ \pi ^0,\mathcal H\}=-\partial_j \pi^j$. Thus we impose $\phi_2=\partial_j\pi^j\approx 0$ as a secondary constraint, which must be added to (\ref{ext_ham_1}) with a corresponding Lagrange multiplier. The total Hamiltonian now reads
\begin{equation}
 \mathcal H=\mathcal H_0+u_1(x) \pi^0(x) +(u_2(x)-A_0)\partial_j\pi^j,
\end{equation}
with
\begin{equation}
\label{classicalhamiltonian}
\mathcal H_0=-\frac{1}{2}M_{js}\pi^j \pi^s-\pi^j M_{ij}G^{0irk}\partial_r A_k+\frac{1}{2}( G^{mnjk}- M_{ir}G^{0ijk} G^{0rmn} )\partial_m A_n \partial_j A_k.
\end{equation}
Now we find $\{ \phi_2,\mathcal H\}=0$, so that the Dirac-Bergmann algorithm ends here and $\phi_1\approx 0$ and $\phi_2\approx 0$ exhaust the constraints. However, $\{\phi_1(t,\vec x),\phi_2(t,\vec y)\}=0$, so that $\phi_1$ and $\phi_2$ are first class constraints, implying that the multipliers $u_1(x)$ and $u_2(x)$ are completely undetermined. The infinitesimal gauge transformations induced by $(\phi_1,\phi_2)$ on the canonical variables $(A_\alpha,\pi^\alpha)$ are
\begin{eqnarray}
\label{gaugetransformations}
\delta A_\alpha (t,\vec x)&=&\int d^3 y\, \epsilon^I(t,\vec y)\{A_\alpha(t,\vec x),\phi_I(t,\vec y)\}=\epsilon^1(t,\vec x)\delta^0_\alpha-\delta^i_\alpha \partial_i \epsilon^2(t,\vec x)\,,\\
\delta \pi^\alpha (t,\vec x)&=&\int d^3 y\, \epsilon^I(t,\vec y)\{\pi^\alpha(t,\vec x),\phi_I(t,\vec y)\}=0,
\end{eqnarray}
with $I=1,2$ and $\epsilon^I(t,\vec x)$ being the infinitesimal parameters of the transformations. Knowledge of these generators of gauge transformations allows us to identify classical observables of the theory as those functionals that are invariant under gauge transformations. Equivalently, observables commute with the constraints $\{O,\phi_{I}\}\approx 0$. In the present case, it can be checked that the electromagnetic inductions
\begin{eqnarray}
D^a &=& -G^{0a0b} F_{0b}-\frac{1}{2} G^{0abk}F_{bk} \\ \nonumber 
&=& -G^{0a0b}\partial_0 A_j-G^{0ab0}\partial_j A_0-G^{0abk}\partial_b A_k\,,\\ 
H_a&=&-\frac{1}{2} \epsilon_{0abc}\left[  G^{bcm0} F_{m0}+\frac{1}{2} G^{bcmn} F_{mn} \right] \\ \nonumber
& =& -\frac{1}{2} \epsilon_{0abc}\left[ G^{bcm0} (\partial_mA_0-\partial_0A_m)+G^{bcmn}\partial_m A_n\right]\,, 
\end{eqnarray}
defined with respect to the chosen foliation of spacetime into spacelike hypersurfaces, indeed commute with the constraints, so that they can be used as observables. Thus we are finally able to write the Hamiltonian (\ref{classicalhamiltonian}) for our system in terms of gauge-invariant observables $D^a$ and $H_a$ as
\begin{equation}
\mathcal H_0=\frac{1}{2} U_{al} D^a D^l+ \frac{1}{2} V^{al} H_a H_l, 
\end{equation}
where the matrices $U$ and $V$ are given as
\begin{eqnarray}
U_{al}&=&-M_{al}+\frac{1}{8} T_{pqjk}G^{0sjk} G^{0tpq}M_{s(l}M_{a)t}\\
V^{al}&=&-\frac{1}{8} \epsilon^{0jk(a} \epsilon^{|0|l)pq} T_{pqjk},
\end{eqnarray}
with $T_{pqjk}$ defined such that 
\begin{equation}
\label{observ}
 (G^{pqmn}-G^{0rpq}G^{0amn}M_{ar})T_{pqtu}=-8 \,\delta^m_{[t} \delta^n_{u]}\,.
\end{equation}
The existence of $T$ is guaranteed due to the invertibility properties of area metrics; indeed it can be written explicitly in terms of the block matrices constituting the area metric tensor, see again appendix \ref{appendixone}.

\section{Gauge fixing and Dirac brackets}\label{sec_dirac}
In order to determine the Dirac brackets associated with our system, one needs to remove the indeterminacy in the Lagrange multipliers by fixing a gauge. This is achieved here by manually imposing two further constraints $\phi_3\approx 0, \phi_4\approx 0$ such that $\textup{det} \{\phi_I(\vec x),\phi_J(\vec y)\} \neq 0$, with $I,J=1,\dots 4$, so that the new set of constraints $\phi_I$ is now of second class.
In our case, the Euler-Lagrange equations for the gauge field $A$ obtained from the action (\ref{areametricaction}) are given by
\begin{equation}
 G^{\alpha \beta \gamma \delta}\,\partial_{\beta}\,\partial_{\delta}\, A_\gamma=0,
\end{equation}
which is conveniently split into one temporal equation
\begin{equation}
 \label{temporalequation}
G^{0a0b}\,\partial_a\,\partial_b A_0+\left [G^{0abc}\partial_a \, \partial_c-G^{0a0b}\partial_0 \,\partial_a\right]A_b  =  0
\end{equation}
and three spatial equations
\begin{equation} 
\label{spatialequation}
\left[ G^{0bla}\,\partial_a\,\partial_b - G^{0l0m}\,\partial_0\,\partial_m \right]A_0  +\left[G^{0l0m}\partial_0^2 -2 G^{0(lm)a}\partial_0 \,\partial_a +G^{lamd} \partial_a\,\partial_d\right]A_m =0\,.
\end{equation}
As the third constraint we impose the Glauber gauge
\begin{equation}
\label{glauber_gauge}
\phi_3=A_0(\vec x)-\int d^3 x'\,G(\vec x,\vec x')G^{0abc}\partial'_a\,\partial'_c A_b(\vec x')\approx 0
\end{equation}
with $-G^{0a0b}\,\partial_a\,\partial_b\,G(\vec x,\vec x')=\delta(\vec x-\vec x' )$, or more explicitly,
\begin{equation}
G(\vec x,\vec x')= -\frac{1}{4\pi \sqrt{-M_{ab}(x^a-x'^a)(x^b-x'^b)}}. 
\end{equation}
The expression under the square root is non-negative ultimately due to the energy distinguishing property (see appendix \ref{appendixone}). Consistency of the gauge (\ref{glauber_gauge}) with the temporal equation (\ref{temporalequation}) requires that the last constraint
\begin{equation}
 \phi_4=G^{0a0b}\,\partial_a A_b\approx 0.
\end{equation}
%\subsection{Restriction to time-reversal and parity symmetric case }
%As we are looking for a non-trivial quantum effect coming from considering a non-metric background, we will restrict our considerations to area metrics $G$ such that $G^{0abc}=\phi \epsilon^{0abc}$, which corresponds to work in a frame in which parity and time-reveral invariance hold \new{still has to be proven that this is the only one solution for preserved time-reveral and parity invariance}. 
In summary, our constraints $\phi_I$ are given by
\begin{equation}
\label{gaugefixedconstraints}
\begin{array}{ll}
\phi_1=\pi^0\approx 0,\quad \quad \quad &\phi_3=A_0(\vec x)-\int d^3 x'\,G(\vec x,\vec x')G^{0abc}\partial'_a\,\partial'_c A_b(\vec x')\approx0,\\
\phi_2=\partial_a \pi^a\approx 0,\quad \quad \quad & \phi_4=G^{0a0b}\,\partial_a\,A_b\approx 0,\\
\end{array}
\end{equation}
and satisfy
\begin{eqnarray}
\{\phi_I(t,\vec x),\phi_J(t,\vec y)\}
&=&\int \dfrac{d^3k}{(2\pi^3)}\left[
\begin{array}{cccc}
0 & 0 & -1 & 0\\
0 & 0 & 0 & -G^{0a0b}\,k_a k_b\\
1 & 0 & 0 & 0\\
0 & G^{0a0b} k_a k_b & 0 &0\\
\end{array}
\right]e^{i \vec k.(\vec x-\vec y)}.
\end{eqnarray}
%\begin{equation}
%\label{constraint_matrix}
%\begin{array}{ccccc}
%\{\phi_I(\vec x),\phi_J(\vec y)\}&=& &\left[
%\begin{array}{cccc}
%0 & 0 & -1 & 0\\
%0 & 0 & 0 & G^{0a0b}\,\partial_a\,\partial_b\\
%1 & 0 & 0 & 0\\
%0 & -G^{0a0b} \partial_a\,\partial_b & 0 &0\\
%\end{array}
%\right]&\delta(\vec x-\vec y)\\
%&=&\int \dfrac{d^3k}{(2\pi^3)}&\left[
%\begin{array}{cccc}
%0 & 0 & -1 & 0\\
%0 & 0 & 0 & -G^{0a0b}\,k_a k_b\\
%1 & 0 & 0 & 0\\
%0 & G^{0a0b} k_a k_b & 0 &0\\
%\end{array}
%\right]&e^{i \vec k.(\vec x-\vec y)}\\
%\end{array},
%\end{equation}
The matrix above $\{\phi_I(t,\vec x),\phi_J(t,\vec y)\}$ is invertible, so that the constraints $\phi_I$ are now of second class and the gauge freedom is gone. Its inverse $(\{\phi(\vec x),\phi(\vec y)\}^{-1})^{IJ}$, defined through
\begin{equation} 
\int\,d^3 y\,\{\phi_I(\vec x),\phi_J(\vec y)\}\, (\{\phi(\vec y),\phi(\vec z)\}^{-1})^{JM}=\delta_I^M\delta(\vec x-\vec z),
\end{equation} 
is simply given as
\begin{equation}
\label{inverse_constraint_matrix}
\{\phi_I(t,\vec x),\phi_J(t,\vec y)\}^{-1}=\int \dfrac{d^3k}{(2\pi^3)}
\left[
\begin{array}{cccc}
0 & 0 & 1 & 0\\
0 & 0 & 0 & \dfrac{1}{G^{0a0b}\,k_a k_b}\\
-1 & 0 & 0 & 0\\
0 & -\dfrac{1}{G^{0a0b} k_a k_b} & 0 &0\\
\end{array}
\right]\,e^{i \vec k.(\vec x-\vec y)}\\.
\end{equation}
Equipped with equation (\ref{inverse_constraint_matrix}) we can now follow Dirac's procedure and replace the standard Poisson bracket $\{,\}$ by the Dirac bracket $\{,\}_D$, which is defined as
\begin{equation}
\label{diracbracket}
\{A(\vec x),B(\vec y)\}_D=\{A(\vec x),B(\vec y)\}-\int\, d^3 z \,d^3 w\,\{A(\vec x),\phi_I(\vec z)\}(\{\phi(\vec z),\phi(\vec w)\}^{-1})^{IJ}\{\phi_J(\vec w),B(\vec y)\}.
\end{equation}
Thus we arrive at the fundamental Dirac brackets of our system, with respect to which the theory must be quantized
\begin{eqnarray}
\label{fuldameltal_dirac_conmutators}\nonumber
\{A_\alpha(t,\vec x),\pi^\beta(t,\vec y)\}_D&=&\int\,\dfrac{d^3 k}{(2\pi)^3}\left[\delta^\beta_\alpha-\delta^0_\alpha \delta^\beta_0-\dfrac{\delta^m_\alpha \delta^\beta_n \, k_a\,k_m\,G^{0a0n}}{G^{0p0q}k_p\,k_q}-\dfrac{\delta^0_\alpha \delta^\beta_b G^{0abc}\,k_a\, k_c}{G^{0p0q}k_p\,k_q}\right]\,e^{i\vec k.(\vec x-\vec y)}\,,\\
\{A_\alpha(t,\vec x),A_\beta(t,\vec y)\}_D&=&0\,,\\\nonumber
\{\pi^\alpha(t,\vec x),\pi^\beta(t,\vec y)\}_D&=&0\,,
 \end{eqnarray}
and the dynamics of the system is simply generated by the Hamilton equations
 \begin{eqnarray}\label{weak_dynamics}
 \partial_t A_\alpha(t,\vec x)&\approx &\int d^3 y\, \{A_\alpha(t,\vec x),\mathcal H_0(\vec y) \}_D\,,\\\nonumber
 \partial_t \pi^\alpha(t,\vec x)&\approx &\int d^3 y\, \{\pi^\alpha(t,\vec x),\mathcal H_0(\vec y) \}_D\,,
 \end{eqnarray}
where, due to the use of Dirac brackets, only  $\mathcal H_0$ is involved.
%\begin{equation}
% \begin{array}{lcl}
%\{A_\alpha(t,\vec x),\pi^\beta(t,\vec y)\}_D&=&\int\,\dfrac{d^3 k}{(2\pi)^3}\left[\delta^\beta_\alpha-\delta^0_\alpha \delta^\beta_0-\delta^m_\alpha \delta^\beta_n \,\dfrac{k_a\,k_m\,G^{0a0n}}{G^{0p0q}k_p\,k_q}\right]\,e^{i\vec k.(\vec x-\vec y)}\\
%\{A_\alpha(t,\vec x),A_\beta(t,\vec y)\}_D&=&0\\
%\{\pi^\alpha(t,\vec x),\pi^\beta(t,\vec y)\}_D&=&0\\
% \end{array}
%\end{equation}

\section{Bihyperbolic area metrics close to Lorentzian metrics}\label{sec_classI}
The preceding Hamiltonian analysis and calculation of Dirac brackets made only implicit use of the requirement that the area metric background be bi-hyperbolic and energy-distinguishing, namely in the abstract constructions underlying the definition of spacetime foliations into spacelike leaves. But now we need to explicitly  solve the field equations (\ref{spatialequation}) with the gauge imposed by (\ref{glauber_gauge}), and this requires to restrict attention to concrete bi-hyperbolic and energy-distinguishing area metric backgrounds. Moreover, for actual calculations it is most convenient to choose a coordinate frame in which the area metric takes a simple normal form. The normal form theory of area metrics in four dimensions has been developed in \cite{schuller2010}, and used in \cite{raetzel2010} to show that the area metric cannot be bi-hyperbolic unless the endomorphism $J$ on the space of two-forms defined through
\begin{equation}
   J_{\gamma \delta}{}^{\alpha \beta} = G^{\gamma \delta \mu \nu} \omega_{\mu \nu \alpha \beta} 
\end{equation}
has a complex eigenvalue structure (Segr\'e type) of the form $[1\bar 1 1 \bar 1 1 \bar 1]$, $[2\bar 2 1 \bar 1]$, $[3 \bar 3]$, $[1 \bar 1 1 \bar 1 1 1]$, $[2 \bar 2 1 1]$, $[1 \bar 1 11 11]$ or $[11 11 11]$. However, four-dimensional area metrics that are induced by a Lorentzian metric automatically lie in the first class, $[1 \bar 1 1 \bar 1 1 \bar 1]$, and moreover the continuous dependence of the eigenvalues of an endomorphism on the components of a representing matrix implies that any area metric in the neighborhood of such a metric-induced area metric is equally of class $[1 \bar 1 1 \bar 1 1 \bar 1]$. Thus area metrics of immediate phenomenological relevance are clearly those of this first class, and it can be shown that by $GL(4)$ frame transformations these can always be brought to the form
\begin{equation}
\label{classInormalform}
G^{[ab][cd]}=
\left[\begin{array}{cccccc}
-\alpha & 0 & 0 & \rho & 0 & 0\\
0 & -\beta & 0 & 0 & \sigma & 0\\
0 & 0 &-\gamma & 0 & 0 & \tau \\
\rho & 0 & 0 & \alpha \,\,\, & 0 & 0\\
0 & \sigma & 0 & 0 & \beta \,\,\, & 0\\
0 & 0 & \tau & 0 & 0 & \gamma \,\,\,\\
\end{array}\right]\qquad \begin{array}{l}\textrm{for real } \rho, \sigma, \tau\, \textrm{ and } \\ \textrm{real positive } \alpha, \,\beta,\,\gamma\,,\end{array}
\end{equation}
where for notational purposes, $G$ is considered here as a bilinear form on the space of two-forms for convenience, and the representing matrix shown above is with respect to the obvious induced basis in the order $[01], [02], [03], [23], [31], [12]$. The positivity of $\alpha\,,\beta\,,\gamma$ follows from our convention that $P$ is positive on the hyperbolicity cone $C$. If and only if the area metric is induced by a Lorentzian metric, do the real scalars assume the values $\alpha=\beta=\gamma=1$ and $\rho=
\sigma=\tau=0$. So any finite (but not too large) deviation from a Lorentzian metric is encoded in these scalars and the frame that brings about this normal form.

It is straightforward to show that if one chooses $\rho=\sigma=\tau$, the polynomial 
\begin{eqnarray}\label{classI_pol}
  P(q) &=& \alpha\beta\gamma (q_0^4 +  q_1^4 + q_2^4 + q_3^4)\nonumber\\
       & & + \alpha(\beta^2+\gamma^2)(q_2^2 q_3^2 - q_0^2 q_1^2)\nonumber\\       
       & & + \beta(\alpha^2+\gamma^2)(q_1^2q_3^2-q_0^2 q_2^2)\nonumber\\
       & & + \gamma(\alpha^2+\beta^2)(q_1^2q_2^2-q_0^2 q_3^2)
 \end{eqnarray} 
associated with an area metric of this class is hyperbolic with respect to $h=L^{-1}(\partial/\partial t)$. This is most efficiently verified in the normal frame by observing that for $h=(1,0,0,0)$, the real symmetric Hankel matrix $H_1(P_{q,h})$ associated with the polynomial $P_{q,h}$ is positive definite for any covector $q$, which implies that $P$ is hyperbolic \cite{gantmacher1959,basu2006}. The dual polynomialm $P^\#$ takes precisely the same shape in the normal form frame employed here, and thus is also seen to be hyperbolic. Finally also the energy-distinguishing property is easily checked.
Finally, note that for area metrics with polynomial (\ref{classI_pol}), we have
\begin{equation}
\label{specialchoice}
G^{0abc} = \rho\, \epsilon^{0abc}
\end{equation}
in this normal form frame, which significantly simplifies the field equations (\ref{temporalequation}) and (\ref{spatialequation}) whose solutions we will now be able to obtain, orthogonalize appropriately, and thus obtain a diagonalization of the Hamiltonian.

It is worth noting that the hyperbolic polynomial (\ref{classI_pol}) only factorizes if at least two of the scalars $\alpha,\, \beta,\,\gamma$ coincide, so that area metrics with a bi-metric dispersion relation merely present a subset of measure zero within the set of area metrics neighboring Lorentzian metrics. Indeed, for the generic case of mutually different scalars, the polynomial $P$ is irreducible. Thus theories trying to account for birefringence in linear electrodynamics by some sort of bi-metric geometry  fail to parametrize almost all relevant geometries near Lorentzian metric ones.     

\section{Diagonalization of the Hamiltonian}\label{sec_diagonalization}
In order to diagonalize the Hamiltonian (\ref{classicalhamiltonian}) for bi-hyperbolic and energy-distinguishing general linear electrodynamics with a higher-order polynomial dispersion relation given by (\ref{classI_pol}), we first need to find the solutions of the classical field equations (\ref{temporalequation}) and (\ref{spatialequation}). After choosing the Glauber gauge (\ref{gaugefixedconstraints}), the first equation is trivially satisfied, and the second one reduced to
\begin{equation}
\label{final_field_eq}
\left[G^{0l0m}\partial_0^2+G^{lamd} \partial_a\,\partial_d\right]A_m(t,\vec x) =0\,,
\end{equation}
due to (\ref{specialchoice}). Moreover, these field equations are completely equivalent to the field equations arising from (\ref{weak_dynamics}). Specifically, we look for plane wave solutions
\begin{equation}
\label{ansatz}
A_a(t,\vec x)=\int \frac{d^3 p}{(2\pi)^3} e^{-i(\omega t+\vec p.\vec x)}f_a(\vec p),
\end{equation}
so that introducing (\ref{ansatz}) into (\ref{final_field_eq}) we observe that the equation
\begin{equation}\label{eigenvalueequation}
\left[G^{0l0m}(\omega)^2+G^{lamd} p_a\,p_d\right]f_m(\vec p) =0
\end{equation}
must be satisfied if (\ref{ansatz}) is indeed a solution. Equation (\ref{eigenvalueequation}) has non-trivial solutions only if \begin{equation}
\label{polyn_def}
\textup{det}\left(G^{0l0m}(\omega)^2+G^{lamd} p_a\,p_d\right)=0\,.
\end{equation}
The non-zero frequencies $\omega$ for which this is the case are precisely the solutions of $P(\omega,\vec p)=0$, compare (\ref{fresnel_det}). From the energy distinguishing condition of an area metric spacetime it follows that this frequencies are non-zero unless $\vec p=0$, and real because of the hyperbolicity of $P$. 
It is then further immediate from (\ref{classI_pol}) that if some (without loss of generality positive) $\omega(\vec{p})$ is a solution for some given $\vec{p}$ in our normal frame, then so is $-\omega(\vec{p})$, and that $\omega(\vec{p})=\omega(-\vec{p})$. Thus we have four non-zero energy solutions $\pm\omega^I(\vec{p})$ labeled by $I=1,2$, two positive and two negative ones, for each spatial momentum $\vec{p}$.  
Therefore any solution of the field equations for the real gauge potential $A$ can be expanded as 
\begin{equation}
\label{generalsolution}
 A_a(t,\vec x)= \sum_{I=1,2} \int_{N^{\textup{smooth}}} \frac{d^3 p}{(2\pi)^3}\left( e^{-i(\omega^I(\vec p) t+\vec p.\vec x)}f^I_a(\vec p)+e^{i(\omega^I(\vec p) t+\vec p.\vec x)}f^{*I}_a(\vec p)\right)\,,
\end{equation}
where strictly speaking, the integral is to be taken only over spatial momenta $\vec{p}$ for which the roots $\omega$ of $P(\omega,\vec{p})$ are non-degenerate, so that the elementary plane wave solutions are linearly independent. However, the set of covectors for which these zeros are degenerate is of measure zero \cite{raetzel2010}, so that this restriction of the integral domain can be technically disregarded. It may be worth emphasizing that the standard appearance of this expansion is somewhat deceptive, since the $\omega^I$ appearing here are solutions of (\ref{polyn_def}), rather than the standard Lorentzian dispersion relation.

Having obtained a basis of solutions of the classical field equations, we now identify an inner product that is preserved under time evolution and positive definite for positive energy solutions. To this end, consider solutions $A_a(\vec p)(t,\vec x)$ and $\tilde A_a(\vec q)(t,\vec x)$ of the field equation for specific spatial covectors $\vec p$ and $\vec q$, respectively. Using the field equation (\ref{final_field_eq}), it can be shown that the continuity equation
\begin{equation}
\label{continuityequation}
 \partial_0\left(G^{0a0b}A_a^*(\vec p)\overleftrightarrow{\partial}_0 \tilde A_b(\vec q)\right)+
\partial_m\left(-G^{a(mn)b} A_a^*(\vec p)\overleftrightarrow{\partial}_n \tilde A_b(\vec q) \right)=0
\end{equation}
is satisfied. This implies that we have a conserved charge $Q$ given by 
\begin{equation}
Q=\int d^3x \,G^{0a0b}A_a^*(\vec p)\overleftrightarrow{\partial}_0 \tilde A_b(\vec q).
\end{equation}
The above defined charge $Q$ can be used to define a scalar product in the space of solutions, which then by definition is conserved under time evolution and is defined as $(A(\vec p),\tilde A(\vec q))=-i\,Q$. It satisfies the following properties
\begin{eqnarray}\label{scalar_product}
\nonumber
(A(\vec p),\lambda \tilde A(\vec q))&=&\lambda(A(\vec p),\tilde A(\vec q))\\
(\lambda A(\vec p),\tilde A(\vec q))&=&\lambda^*(A(\vec p),\tilde A(\vec q))\\\nonumber
(A(\vec p),\tilde A(\vec q))&=&( A(\vec p),\tilde A(\vec q))^*=-( A^*(\vec p),\tilde A^*(\vec q)).
\end{eqnarray} 
Hence, if we define for our different frequency solutions
\begin{equation}
 F^I_{a}(\vec p)(t,\vec x)=e^{-i(\omega^I(\vec p) t+\vec p.\vec x)}f^I_a(\vec p),
\end{equation}
we find that $(F^I(\vec p),F^{*J}(\vec q))=0$ and
\begin{equation}
\label{positivesolutions}
(F^I(\vec p),F^{J}(\vec q))=-(F^{*I}(\vec p),F^{*J}(\vec q))=-2\omega^I(\vec p) G^{0a0b} f^{I*}_a(\vec p) f^{I}_a(\vec p) \delta^{IJ}\delta(\vec p-\vec q).
\end{equation}
In the derivation of the above results we used charge conservation to find that for $I\neq J$
\begin{equation}
\label{ortho_id}
 G^{0a0b} f^{I*}_a(\vec p)f^{J^*}_b(-\vec p)=G^{0a0b} f^{I*}_a(\vec p)f^{J}_b(\vec p)=0.
\end{equation}
%The last expressions are very useful when quantizing the theory. 
Moreover, since $G^{0a0b}$ is negative definite due to (\ref{classInormalform}), equation (\ref{positivesolutions}) shows that the positive energy solutions can be positively normalized, implying in turn that the negative energy solutions are negatively normalized. This indefiniteness of the scalar product is responsible for creation and anhilation processes. Choosing, without loss of generality, $f^{I}_a(\vec p)=\frac{a^I_a(\vec p)}{\sqrt{2\omega^I(\vec p)}}$, we finally have
\begin{equation}
(F^I(\vec p),F^{J}(\vec q))=-(F^{*I}(\vec p),F^{*J}(\vec q))=- G^{0a0b} a^{I*}_a(\vec p) a^{I}_a(\vec p) \delta^{IJ}\delta(\vec p-\vec q),
\end{equation}
and our general solution reads
\begin{eqnarray}
\label{finalsolution}
 A_a(t,\vec x)&=&\sum_{I=1,2} \int \frac{d^3 p}{(2\pi)^3}\frac{1}{\sqrt{2\omega^I(\vec p)}}\left( e^{-i(\omega^I(\vec p) t+\vec p.\vec x)}a^I_a(\vec p)+e^{i(\omega^I(\vec p) t+\vec p.\vec x)}a^{*I}_a(\vec p)\right).
\end{eqnarray}
Now that we have the general solution (\ref{finalsolution}), we can use it to write the Hamiltonian evaluated at a solution in diagonal form,
\begin{eqnarray}
\label{class_Ham}
{H}_0=\int d^3x\,\mathcal H_0(\vec x)&=&-\frac{1}{2}\int d^3 x G^{0a0b}\left( \partial_0 A_a \partial_0 A_b-A_a\partial_0^2 A_b\right)\\\nonumber
&=&\frac{1}{2}\sum_{I,J} \int \frac{d^3 p}{(2\pi)^3}\frac{d^3 q}{(2\pi)^3} \omega^J(\vec p)\left[ (F^I(\vec p),F^J(\vec q))+(F^J(\vec p),F^I(\vec q))\right]\\\nonumber
&=&-\frac{1}{2}\sum_{I=1,2} \int \frac{d^3 p}{(2\pi)^3} \omega^I(\vec p) G^{0a0b}\left[ a^{*I}_a(\vec p)a^I_b(\vec p))+a^I_a(\vec p) a^{*I}_b(\vec p))\right].
\end{eqnarray}
The last expression shows that the classical Hamiltonian is positive  because $G^{0a0b}$ is negative definite.

\section{Quantization}
\label{sec_quantization}
Equipped with the results developed so far, we are now ready to quantize the electromagnetic field. First,  notice that if we multiply equation (\ref{eigenvalueequation}) by $p_l$ then the amplitude eigenvectors $a^I_b(\vec p)$ satisfy 
\begin{equation}
\label{orthogonality}
G^{0a0b}p_a a^I_b(\vec p)=0,
\end{equation}
such that the constraints $G^{0a0b}\partial_a A_b\approx 0$ and $\partial_a \pi^a\approx 0$ are satisfied. Now it can be shown \cite{raetzel2010} that for almost all spatial momenta $\vec{p}$, the two associated positive energies do not coincide, $\omega^{I=1}(\vec{p}) \neq \omega^{I=2}(\vec{p})$, so that the covectors $a^{I=1}_b(\vec p)$ and $a^{I=2}_b(\vec{p})$ are determined up to scale, linearly independent and thus forming a basis for the space of all purely spatial covectors $v$ for which $G^{0a0b}p_a v_b=0$, and diagonalize the Hamiltonian (\ref{class_Ham}). Thus the only one freedom we have is a choice of normalization, which we choose such that any solution $a^I_b(\vec p)$ is expressed as $a^I_b(\vec p)=a^I(\vec p) \epsilon^I_b(\vec p)$ with the covectors $\epsilon^I_b(\vec p)$ normalized with respect to our scalar product, i.e.,
\begin{equation}
-G^{0a0b} \epsilon^{I^*}_a(\vec p) \epsilon^I_b(\vec p)=1,
\end{equation}
where there is no summation over $I$. Furthermore, $p_a$ and any $a^I_b(\vec p)$ are clearly linearly independent, such that the set of covectors
\begin{equation}
\left\lbrace \epsilon^{I=1}_b(\vec p),\epsilon^{I=2}_b(\vec p),\frac{\vec p}{\sqrt{-G^{0a0b}p_a p_b}}\right\rbrace 
\end{equation}
constitute a basis for $V$, which is orthonormalized with respect to the scalar product (\ref{scalar_product}). Hence, they satisfy the completeness relation
\begin{equation}
\label{completeness}
-G^{0i0j}\sum_{I=1,2}\epsilon^{I*}_j(\vec p) \epsilon^{I}_b(\vec p)=\delta^i_b-\frac{p_m p_b G^{0m0i}}{G^{0r0s}p_r p_s}.
\end{equation}
Notice that the normalized covectors $\epsilon^I_b(\vec p)$ satisfy the orthogonality identities (\ref{ortho_id}). Now the general solution (\ref{finalsolution}) takes the form
\begin{eqnarray}
\label{finalsolution2}
 A_a(t,\vec x)&=&\sum_{I=1,2} \int \frac{d^3 p}{(2\pi)^3}\frac{1}{\sqrt{2\omega^I(\vec p)}}\left( e^{-i(\omega^I(\vec p) t+\vec p.\vec x)}a^I(\vec p)\epsilon^I_a(\vec p)+e^{i(\omega^I(\vec p) t+\vec p.\vec x)}a^{I*}(\vec p)\epsilon^{*I}_a(\vec p)\right)\,,
\end{eqnarray}
where the coefficients $a^{I}(\vec p)$ correspond to the amplitudes of the solutions and depend on the initial values that one considers for a specific problem in the classical approach.  At the quantum level, these amplitudes are precisely the mathematical objects that should be promoted to operators, such that the corresponding quantum field reads
\begin{eqnarray}
\label{quantumsolution}
 \hat A_a(t,\vec x)&=&\sum_{I=1,2} \int \frac{d^3 p}{(2\pi)^3}\frac{1}{\sqrt{2\omega^I(\vec p)}}\left( e^{-i(\omega^I(\vec p) t+\vec p.\vec x)}\hat a^I(\vec p)\epsilon^I_a(\vec p)+e^{i(\omega^I(\vec p) t+\vec p.\vec x)}\hat a^{I \dagger}(\vec p)\epsilon^{*I}_a(\vec p)\right).
\end{eqnarray}
Using this quantum solution and the expressions for the energy and spatial momentum (which can be obtained by calculating the energy-momentum tensor) we find that the quantum Hamiltonian and quantum spatial momentum operators are given by
\begin{eqnarray}
\label{quantumhamiltonian}
{\hat H}_0&=&\frac{1}{2}\sum_{I=1,2} \int \frac{d^3 p}{(2\pi)^3} \omega^I(\vec p) \left[ \hat a^{I}(\vec p)\hat a^{I\dagger}(\vec p))+\hat a^{I\dagger}(\vec p)\hat a^{I}(\vec p))\right],\\
{\hat P}_i&=&\frac{1}{2}\sum_{I=1,2} \int \frac{d^3 p}{(2\pi)^3}\, p_i\, \left[ \hat a^{I}(\vec p)\hat a^{I\dagger}(\vec p))+\hat a^{I\dagger}(\vec p)\hat a^{I}(\vec p))\right].
\end{eqnarray}
Hence, if we identify the operators $\hat a^I(\vec p),\,\hat a^{I\dagger}(\vec p)$ with annihilation and creation operators respectively, a condition for the Hamiltonian to be positive definite is that these operators obey the bosonic commutation relations
\begin{eqnarray}
\label{crea_anh_comm}
[\hat a^I(\vec p), \hat a^{J\dagger}(\vec q)]&=&(2\pi)^3\delta^{IJ}\delta(\vec p-\vec q)\,,\\\nonumber
[\hat a^I(\vec p),\, \hat a^{J}(\vec q)\,] &=& [\hat a^{I\dagger}(\vec p), \hat a^{J\dagger}(\vec q)]=0.
\end{eqnarray}
Hence, the quantum Hamiltonian operator can be written as
\begin{eqnarray}
\label{quantumhamiltonian_normal}
{\hat H}_0&=&\sum_{I=1,2} \int \frac{d^3 p}{(2\pi)^3} \omega^I(\vec p)  \hat a^{I\dagger}(\vec p) \hat a^{I}(\vec p)) +\sum_{I=1,2} \frac{1}{2}\int d^3 p \,\omega^I(\vec p)\delta(0)\,,
\end{eqnarray}
from which expression we identify the energy of the electromagnetic vacuum, which was calculated here for plane wave solutions without any boundary conditions, as
\begin{equation}
\label{zeropointenergy}
E_{\textrm{vac}}(\textrm{no boundaries})=\sum_{I=1,2} \frac{1}{2}\int d^3 p \,\omega^I(\vec p)\delta(0)\,.
\end{equation}
In the next section, we will calculate how this expression changes if one imposes boundary conditions. Finally, by using the completeness relation (\ref{completeness}) one confirms that
\begin{equation}
\left[ \hat A_a(t,\vec x), \hat \pi^b(t,\vec y)\right] = i\int\,\dfrac{d^3 p}{(2\pi)^3}\left[  \delta^b_a-\frac{p_m p_a G^{0m0b}}{G^{0r0s}p_r p_s} \right]\,e^{i\vec p.(\vec x-\vec y)},
 \end{equation}
which shows the consistency of the quantization procedure with the Dirac brackets (\ref{fuldameltal_dirac_conmutators})\new{, since the latter reduce to the above form due to (\ref{specialchoice}).}

\section{Application: Casimir effect in a birefringent linear optical medium}
\label{sec_casimir}
\new{The Hamiltonian (\ref{quantumhamiltonian_normal}) shows that the quantization of general linear electrodynamics leads to a modified quantum vacuum compared to standard non-birefringent Maxwell theory. In fact, local physical phenomena which only depend on the quantum vacuum can be used to test and bound the non-metricity of spacetime. In this section we  analyze one such phenomenon, namely the Casimir effect; similar studies can be conducted for the Unruh effect and spontaneous emission.

The Casimir effect \cite{casimir1948} arises because of the energy cost incurred by imposing boundary conditions on the electromagnetic field strength. Physically, such boundary conditions arise for instance by introducing perfectly conducting metal plates into the spacetime. For two infinitely extended plates parallel to the 1-2-plane, and this is the configuration we will study here for general linear electrodynamics, the electromagnetic field strength must satisfy the boundary conditions \begin{equation}\label{platebdy}
   \left.F_{01}\right|_{\textrm{plates}} = \left.F_{02}\right|_{\textrm{plates}}=\left.F_{12}\right|_{\textrm{plates}} =0
\end{equation}
everywhere on either plate; this follows, by Stokes' theorem and thus independent of the geometric background, from the physical assumption that the plates are ideal conductors inside of which the field strength must vanish.

Now the key point is that having, or not having, boundary conditions for the vacuum amounts to an energy difference, the so-called Casimir energy
\begin{equation}
  E_\textrm{Casimir} = E_\textrm{vac}(\textrm{plate boundaries}) - E_\textrm{vac}(\textrm{no boundaries})\,.
\end{equation}
But both energies on the right hand side diverge and need to be regularized such that their difference is independent of the regulator.
This is most easily achieved by first considering boundary conditions analogous to (\ref{platebdy}), but for all six faces of a finite rectangular box with faces parallel to the coordinate planes, and separated by coordinate distances  $L_1, L_2, L_3$. In a second step we will then push all faces a very large coordinate distance $L$ apart in order to obtain an expression for $E_\textrm{vac}(\textrm{no boundaries})$ regularized by $L$, and similarly push all but two faces in order to obtain a corresponding regularized expression for $E_\textrm{vac}(\textrm{plate boundaries})$. The difference of these two regulated quantities will indeed turn out to be finite per unit area and be independent of the regulator $L$.

Now more precisely, a basis of solutions of general linear electrodynamics satisfying the box boundary conditions is labeled by a triple $(n_1,n_2,n_3)$ of non-negative integers and a polarization $I=1,2$ and takes the form
\begin{eqnarray}
\label{metric_solution}\nonumber
A_x(\vec x)&=&a^I_x(n_1,n_2,n_3) \, \textup{cos}(n_1 \pi \frac{x}{L_1})\, \textup{sin}(n_2 \pi \frac{y}{L_2})\, \textup{sin}(n_3 \pi \frac{z}{L_3})\,,\\
A_y(\vec x)&=&a^I_y(n_1,n_2,n_3) \, \textup{sin}(n_1 \pi \frac{x}{L_1}) \, \textup{cos}(n_2 \pi \frac{y}{L_2}) \, \textup{sin}(n_3 \pi \frac{z}{L_3})\,,\\\nonumber
A_z(\vec x)&=&a^I_z(n_1,n_2,n_3) \, \textup{sin}(n_1 \pi \frac{x}{L_1}) \, \textup{sin}(n_2 \pi \frac{y}{L_2})\, \textup{cos}(n_3 \pi \frac{z}{L_3})\,,
\end{eqnarray}
where the $a^I_m(n_1,n_2,n_3)$ are solutions to equation (\ref{eigenvalueequation}) for $\omega^I(n_1 \pi/L_1, n_2 \pi/L_2, n_3 \pi/L_3)$, which always exist if the dispersion relation is bi-hyperbolic and energy distinguishing. The vacuum energy in the presence of the box boundary conditons is thus given by the discrete sum
\begin{equation}
\label{ape_discretized}
E_{vac}(\textrm{box boundaries})=\frac{1}{2}\sum_{\vec n=0}^{\infty} \sum_{I=1,2}\omega^I(\pi \frac{n_1}{L_1}, \pi \frac{n_2}{L_2}, \pi\frac{n_3}{L_3}).
\end{equation}
Removing appropriate faces to a coordinate distance $L$ one finds from this, in the very large $L$ limit, the $L$-regularized expression for the vacuum energy without boundary conditions
\begin{equation}
\label{ape_continuum}
E^L_{vac}(\textrm{no boundaries})= \frac{L^3}{2 \pi^3}\sum_{I=1,2}\int_{0}^{\infty}d^3 p\, \omega^I(\vec p),
\end{equation}
and the $L$-regularized expression for the vacuum energy in the presence of two plates parallel to the 1-2-plane and separated by a coordinate distance $d$
\begin{equation}
E^L_{\textup{vac}}(\textrm{plate boundaries})=\frac{L^2}{2\pi^2} \sum_{I=1,2} \sum_{n'}\int_0 ^\infty dp_x dp_y \,\omega^I\left(p_x^2,p_y^2,(\frac{n\pi}{d})^2\right)\,,
\end{equation} 
where the prime in the summation symbol $n$ means that a factor $1/2$ should be inserted if this integer is zero, for then we have just one independent polarization. Hence we find for the physical vacuum Casimir energy $U(d)=(E_\textrm{vac}(\textrm{plate boundaries}) - E_\textrm{vac}(\textrm{no boundaries}))/L^2$ per unit area 
\begin{equation}
\label{casimir_energy}
U(d) = \frac{1}{2\pi^2} \sum_{I=1,2} \left[ \sum_{n'}\int_0^\infty dp_x dp_y \,\omega^I\left(p_x^2,p_y^2,(\frac{n\pi}{d})^2\right)-\frac{d}{\pi}   \int_0^\infty dp_x dp_y dp_z\,\omega^I(p_x^2,p_y^2,p_z^2)\right].
\end{equation}
In principle, the execution of the above integrals can proceed as in the standard case. However, with the frequencies $\omega^I$ now being solutions to a quartic, rather than quadratic, dispersion relation, these integrals are much harder particularly due to the absence of rotational invariance.
 Fortunately, the fact that contributions from the two different polarizations $I=1,2$ are simply added in the above expression allows for an analytic study of the case where the polynomial $P$ is reducible. In terms of the scalars $\alpha, \beta, \gamma, \rho$ defining the area metric in a normal form frame, this is the case if and only if two of the scalars $\alpha, \beta, \gamma$ coincide, and we may take $\alpha=\beta$, for instance. Even in this simplest of non-trivial cases, the Casimir energy crucially depends on the birefringence properties of the underlying general linear electrodynamics. More precisely, the polynomial in (\ref{classI_pol}) factorizes into two Lorentzian metrics,
\begin{equation}
\label{classIbimetricpolynomial}
 P(p)=\alpha(\alpha p_0^2-\alpha p_3^2-\gamma (p_1^2+p_2^2))(\gamma p_0^2-\gamma p_3^2-\alpha (p_1^2+p_2^2))\,,
\end{equation}
so that we immediately obtain the positive energy solutions
\begin{equation}
\label{bimetric_energies}
\omega^{I=1}=\left[\frac{1}{\alpha}\left(\alpha p_3^2+\gamma p_2^2 + \gamma p_2^2\right)\right]^{1/2}\qquad \textrm{and}\qquad
\omega^{I=2}=\left[\frac{1}{\gamma}\left( \gamma p_3^2+\alpha p_1^2 + \alpha p_2^2\right)\right]^{1/2}\,,
\end{equation}
turning (\ref{casimir_energy}) into a sum of integrals as they appear in the standard Casimir problem on a Lorentzian background. Thus from here on the standard calculation of the Casimir effect \cite{milonni1994} can be followed for each of these integrals separately, and one finally obtains the Casimir energy (\ref{casimir_energy})   
\begin{equation}
U(d)=-\frac{1}{2}\left( \frac{\alpha}{\gamma}+\frac{\gamma}{\alpha} \right) \frac{\pi^2}{720 d^3}\,.
\end{equation}
This energy difference of course results in a Casimir force 
\begin{equation}
\label{casimir_force}
 F(d)=-U'(d)=-\frac{1}{2}\left( \frac{\alpha}{\gamma}+\frac{\gamma}{\alpha} \right) \frac{\pi^2}{240 d^4}
\end{equation}
between the plates. The standard Casimir force is recovered if and only if $\alpha=\beta=\gamma$, and irrespective of the value of the scalar $\rho$. This in turn is equivalent to the absence of classical bi-refringence \cite{favaro2010}. Note that the amplification of any bi-refringence is limited only by the technological constraint of how small the separation $d$ between the plates can be made in any realistic set-up. In contrast to classical bi-refringence tests, which  usually require accumulative effects over large distances (with all the uncertainties present in such non-local measurements), one sees here that the Casmir force allows for a detection of bi-refringence by way of a highly local measurement.  
Conversely, of course, experimental measurements of the Casimir force agreeing with the standard prediction within the given technological constraints can be used to put stringent bounds on the non-metricity of the spacetime region where the measurement is conducted.}

\section{Conclusions}
The canonical quantization of general linear electrodynamics, as undertaken in this article, required the solution of several, and in themselves challenging, questions. 

First, from the classical field theory point of view, it had to be clarified which general linear electrodynamics are predictive on the one hand and physically interpretable in terms of quantities measurable by observers on the other hand. The answer to both questions is encoded in the polynomial dispersion relation of the field theory, and amounts to the simple algebraic conditions that the latter be bi-hyperbolic and energy-distinguishing. Further down the road, these conditions turned out to be crucial in ensuring the existence of a Glauber gauge, which allowed to define a time-conserved scalar product in the space of classical solutions, on which all further developments were based.

Second, and closely related, is the construction of a Hamiltonian formulation of
general linear electrodynamics. The causal structure encoded in the higher-order polynomial dispersion relation of this theory required a revision of the construction of suitable spacetime foliations that underlie a Hamiltonian formulation. The key point here was that the leaves of the foliation must be such that initial data provided on them must be causally evolved by the field equations and at the same time be accessible to observers. It turned out that bi-hyperbolic and energy-distinguishing area metric manifolds provide precisely the structure to ensure both, and ultimately render the classical Hamiltonian positive.    

Third, the quantum Hamiltonian operator is positive definite. For a theory with a higher-order polynomial dispersion relation this is far from trivial, and again only due to bi-hyperbolicity and the energy-distinguishing property. The positive definiteness of the quantum Hamiltonian operator is inherited from the positivity of the classical Hamiltonian because the positive energy solutions have positive norm with respect to the scalar product identified before. This is of course synonymous with the stability of the quantum vacuum, and thus the physical relevance of the Casmir effect we derived from it. 

The wider lesson learnt from our study consists in this being a prototypical, and rather non-trivial example for the quantization of a field theory with a modified dispersion relation. Such theories are discussed extensively throughout the literature with a number of motivations, but usually disregarding the fundamental consistency conditions that were instrumental in this work. In particular, the classically inevitable condition that the dispersion relation be given by a bi-hyperbolic and energy distinguishing polynomial proved inevitable also at virtually every step of the quantization process. 

Actual calculations were made tractable by employing the fact that the dispersion relation of general linear electrodynamics is ultimately determined by a fourth rank area metric tensor for which a complete algebraic classification and associated normal forms are available for the phenomenologically directly relevant case of four spacetime dimensions. This normal form theory was also used to ensure that the birefringent optical backgrounds for which we calculated the Casimir effect (and which owe their physical relevance to their parametrizing the neighborhood of non-birefringent optical media) indeed are bi-hyperbolic and energy-distinguishing. While it is possible to directly exclude 16 out of a total of 23 algebraic classes of four-dimensional area metrics as admissible spacetime structures, a complete and simple characterization of all area metric manifolds that {\it are} bi-hyperbolic and energy-distinguishing however remains an open problem. The high interest that would attach to a comprehensive solution of this problem is clearly underlined by the pivotal role we saw these conditions to play for the classical and quantum theory alike. 

Another open, albeit well-defined, problem is the coupling of fermions to general linear electrodynamics. The issue is the very definition of spinors in the presence of a higher-degree polynomial dispersion relation, rather than one given by a Lorentzian metric. For rather than satisfying the standard binary Dirac algebra, generalized Dirac matrices that intertwine spacetime and spinor indices must now satisfy a quarternary algebra determined by the fourth-degree polynomial associated with a four-dimensional area metric spacetime structure. Even employing the normal form theory, representations of this quarternary algebra appear hard to find in any other but the case of a reducible dispersion relation satisfying the relevant conditions (which then leads to a sixteen-dimensional spinor representation with an associated refined Dirac equation for this special bi-metric case). Once a representation in the general case is obtained, the canonical quantization can proceed exactly along the now clearly defined path for such theories, and complete a full theory of general linear quantum electrodynamics including charges.      

Concluding, we see that the results of this article open up the arena for comprehensive, and above all conceptually watertight, studies of quantum effects brought about by birefringence. Indeed, beyond the Casimir force we calculated here explicitly, any other effect rooting in the quantum vacuum of electrodynamics can be directly calculated now on the basis of the technical findings of this paper. This includes for instance the Unruh effect or the spontaneous emission of photons from quantized point particles. Once spinor fields are included, the range of effects of course extends to the full spectrum of processes discussed in standard quantum electrodynamics with charged fermions. 
Far from being merely academic musings, however interesting, these findings are of immediate relevance to physicists with interests ranging from fundamental theory to material science. Indeed, while on the one hand directly testable in birefringent optical media in laboratory experiments \footnote{For not truly continuous optical media, such as any material in the laboratory, the calculations made in this paper will only apply to the same approximation to which the medium can be modeled as an area metric spacetime. This will be the case for wavelengths that are well above the average distance between the atoms constituting the material but also well below wavelengths of the order of the separation between the plates. More sophisticated calculations taking into account these cut-offs, as well as finitely extended plates or different geometric configurations should now be feasible for the interested specialist, based on the results derived in this paper.}, the constructions of this paper on the other hand also put phenomenological studies of modified dispersion relations \cite{ling2006,barcelo2007,rinaldi2007,garattini2010,bazo2009,gregg2009,sindoni2008,sindoni2009,sindoni2009-2,chang2008,laemmerzahl2009,laemmerzahl2005,liberati2001,perlick2010,gibbons2007,lukierski1995,glikman2001,girelli2007} (or, equivalently, Lorentz-violating spacetime structures \cite{kostelecky1989,gambini1998,alfaro1999,sudarsky2002,myers2003,magueijo2002,mavromatos2010,bojowald2004,jacobson2006,hossenfelder2010}), as they now abound in the literature, on a solid theoretical footing.

\appendix
\Section{area metrics and hyperbolicity}
\label{appendixone}
\new{An area metric in four dimensions takes the following block matrix Petrov form \cite{schuller2010}}
\begin{equation}
 G^{[\alpha\beta][\gamma\delta]}=\left[ 
\begin{array}{cc}
M&K\\
K^T&N
\end{array}
\right],
\end{equation}
where the antisymmetric index pairs $[01],[02],[03],[23],[31],[12]$ label, in this order, the basis in which the matrix is given. The matrices $M,K,N$ are $3\times 3$ matrices \new{are related to the area metric through}
\begin{eqnarray}
M^{ab}&=&G^{0a0b}\,,\\\nonumber
K^a_b&=&\frac{1}{2}\epsilon_{0bmn}G^{0amn}\,,\\\nonumber
N_{ab}&=&\frac{1}{4}\epsilon_{0amn}\epsilon_{0bpq}G^{mnpq},
\end{eqnarray}
\new{where $\epsilon$ is the totally antisymmetric} Levi-Civita symbol. If $M$ in the expression above is invertible, \new{which at the end of this section we will see to be the case} if the correspondig area metric leads to well-posed field equations, then $\det(N-K^T M K)\neq 0$. This ensures the existence of the object $T_{pqtu}$ defined in (\ref{observ}), which can be shown to be explicitly given by
\begin{equation}
T_{abtu}=-2\epsilon_{tuf} \epsilon_{abm} \left( (N-K^T M K)^{-1}\right)^{fm}.
\end{equation}
The principal symbol of the linear field equations governing the dynamics for the electromagnetic field on area metric backgrounds was found in \cite{schuller2010} as the determinant of the $6 \times 6$ matrix
\begin{equation}
\label{matrixareametric}
A^\alpha p_\alpha=
\left[ \begin{array}{cc}
G^{0m0n}p_0-2G^{0(mn)a}p_a&\quad -\frac{1}{2} \epsilon_{nef}G^{efma}p_a\\
\epsilon^{mna} p_a & \quad \delta^n_m p_0
\end{array}\right].
\end{equation}
Using that for any $n\times n$ matrices $A,B,C,D$
\begin{equation}
 \textup{det}\left[\begin{array}{cc}
A& B\\ C & D \end{array}
\right]=\textup{det}(AD-BC)\quad \textup{if} \quad CD=DC,
\end{equation}
we can write the determinant of (\ref{matrixareametric}) as the determinant of a $3\times 3$ matrix as 
\begin{equation}
\label{fresnel_det}
 \textup{det}(A^\alpha p_\alpha)=\textup{det}\left(G^{0m0n}p_0^2-2G^{0(mn)a}p_a p_0+G^{manu}p_a p_u\right)=-p_0^2 P(p)\,,
\end{equation}
where $P$ is precisely the polynomial given in equation (\ref{fresnel}).
% for the case considered there, so that the above determinant already has the i%nformation of the gauge that one should take to study the Hamiltonian formulation. 
Using (\ref{fresnel_det}), the polynomial \new{$P$} can now be expressed in terms of the constitutive matrices $M,K,N$. After calculation one finds
\begin{equation}
 P_G(p_0,\vec p)= a \,p_0^4+b(\vec p)\, p_0^3+c(\vec p)\, p_0^2+d(\vec p)\, p_0+e(\vec p),
\end{equation}
with coefficients
\begin{eqnarray}\label{generalcoefficients}\nonumber
 a&=&-\textup{det}(M^{ab})\,,\\ \nonumber
b(\vec p)&=&\epsilon_{abc}\,G_M^{aclm}\,K^b_l p_m\,,\\
c(\vec p)&=&-\left( N_{rf} \,G_M^{lrmf}+\epsilon_{abc} \epsilon^{tmr}K^b_t K^a_r M^{cl}+2K^e_a K^{[m}_e M^{a]l}+2K^l_a K^{[a}_e M^{m]e}\right)p_l p_m\,,\\ \nonumber
d(\vec p)&=&-2 \epsilon^{ren}\left( 2N_{ea}K^{[m}_r M^{a]l}-K^m_r K^l_a K^a_e\right)p_l p_m p_n\,,\\\nonumber
e(\vec p)&=&-N_{bf} \epsilon^{bmr} \epsilon^{fns}\left( \frac{1}{2} M^{uv}N_{rs}-K^u_r K^v_s\right)p_u p_m p_v p_n,
\end{eqnarray}
where $G^{ambn}_M=M^{ab} M^{mn}-M^{an} M^{mb}$. Using this expression, we find that a necessary condition for the well-posedness of the initial value problem is that the matrix $M^{ab}$ is invertible. Moreover, for an observer's frame, in which the coefficient $b(\vec p)=0$ as explained in \cite{raetzel2010}, the \new{} matrices $K$ and $M$ satisfy
\begin{equation}
K^{[m}_f M^{n]f}=0.
\end{equation}
In this case, equations (\ref{generalcoefficients}) reduce to 
\begin{eqnarray}\label{generalcoefficients_observer}\nonumber
 a&=&-\textup{det}(M^{ab})\,,\\ \nonumber
b(\vec p)&=&0\,,\\
c(\vec p)&=&-\left( N_{rf} \,G_M^{lrmf}+\epsilon_{abc} \epsilon^{tmr}K^b_t K^a_r M^{cl}+2K^e_a K^{[m}_e M^{a]l}\right)p_l p_m\,,\\ \nonumber
d(\vec p)&=&-2 \epsilon^{ren}\left( 2N_{ea}K^{[m}_r M^{a]l}-K^m_r K^l_a K^a_e\right)p_l p_m p_n\,,\\\nonumber
e(\vec p)&=&-N_{bf} \epsilon^{bmr} \epsilon^{fns}\left( \frac{1}{2} M^{uv}N_{rs}-K^u_r K^v_s\right)p_u p_m p_v p_n.
\end{eqnarray}
For area metrics for which there exists a frame such that $G^{0abc}=\rho \epsilon^{0abc}$, such as those considered from section (\ref{sec_classI}) onwards, or equivalently $K^a_b=\phi \delta^a_b$, the polynomial $P_G(p)$ is further reduced to
\begin{eqnarray}
\label{pol_simplified}
 P_G(p_0,\vec p)&=&-p_0^4\textup{det}(M)-p_0^2 (N_{mn}G_{M}^{imjn})p_i p_j-\frac{1}{2}(M^{ij}p_ip_j)(G^{bksl}N_{bs}p_kp_l)\\\nonumber
&=&-p_0^4\textup{det}(M)-p_0^2(N_{mn}G_{M}^{imjn})p_i p_j-\textup{det}(N)(M^{ij}p_ip_j)(N^{-1})^{kl}p_kp_l.
\end{eqnarray}
From the energy distinguishing property of the area metric spacetimes considered here it follows that  $P(p_0,\vec p)=0$ does not have any solutions $p_0=0$ unless $\vec p=0$. But then the matrix  $M^{ab}$ must be of definite signature.  For suppose that this is not the case, then one could find $\vec p \neq 0$ such that $M^{ij}p_ip_j=0$. That would imply extra zero solutions for $p_0$, in contradiction to the energy distinguishing condition. The same holds for the matrix $N$. Thus, without loss of generality, we assume that $M$ is negative definite; then using Descarte's rule of signs, hyperbolicity of $P$  implies that $N$ must be positive definite. The \new{opposite} definiteness of $M$ and $N$ can be shown to be also sufficient for the hyperbolicity of (\ref{pol_simplified}). \new{This} is indeed the case for class I area \new{metrics} (\ref{classInormalform}) with $\rho=\sigma=\tau$. 
\bibliographystyle{ieeetr}
\bibliography{dispersion}

\end{document}